\documentstyle[preprint,aps,psfig]{revtex}
\begin{document}
\title{
Pulse driven switching in one-dimensional  nonlinear photonic band gap 
materials:
a numerical study 
}
\author{
E.~Lidorikis \cite{lido} and C.~M.~Soukoulis
}
\address{
Ames Laboratory-U.S. DOE and Department of Physics and Astronomy,   
Iowa State University, Ames, Iowa 50011 \\
}
\author{\parbox[t]{5.5in}{\small
We examine numerically the time-dependent properties of nonlinear
bistable multilayer structures for constant wave illumination. 
We find that our system exhibits both steady-state and 
 self-pulsing solutions. In the steady
state regime, we examine the dynamics of driving the system between
different transmission states by injecting pulses, and we find the
optimal pulse parameters. We repeat this work for the case of a linear
periodic  system with a nonlinear impurity layer. 
 }}
\maketitle
\normalsize 

\section{INTRODUCTION}

Nonlinear dielectric materials \cite{newell_book} 
exhibiting a bistable response
to intense radiation are  key elements for an all-optical digital
technology.  For certain input optical powers
there may exist  two distinct transmission branches forming
a hysteresis loop, which incorporates a history dependence 
in the system's response. Exciting applications involve 
 optical switches, logic gates, set-reset fast memory elements etc 
\cite{gibbs_book}. 
Much interest has been given lately to periodic nonlinear structures
\cite{winful},
in which because of the distributed feedback mechanism, the nonlinear
effect is greatly enhanced. In the low intensity limit, these
structures are just Bragg reflectors characterized by high transmission
bands separated by photonic band gaps \cite{soukoulis}. 
For high intensities and 
frequencies inside the transmission band, bistability results from the 
modulation of transmission by an intensity-dependent phase shift. 
For frequencies inside the gap bistability originates from gap soliton
formation \cite{chen}, which can lead to much lower switching thresholds 
\cite{desterke2}.

The response of nonlinear periodic structures illuminated by a 
constant wave (CW) with a frequency inside the photonic gap is
generally separated into three regimes: i) steady state response via
stationary gap soliton formation, ii) self-pulsing via excitation 
of solitary waves, and  iii) chaotic.  Much theoretical work has
been done  for  systems with a weak sinusoidal 
refractive index modulation and uniform nonlinearity 
\cite{winful,mills,desterke,wabnitz,john}, or deep modulation 
multilayered systems \cite{chen,dowling,tran}, 
as well as experimental \cite{sankey,eggleton}. 
One case of interest
is when the system is illuminated by a CW bias and switching between
different transmission states is achieved by means of external pulses. 
Such switching has already been 
demonstrated experimentally for various kinds of nonlinearities 
\cite{sankey,tarng,cada2}, 
but to our knowledge, a detailed study
of the dynamics, the optimal pulse parameters and the stability 
under phase variations during injection, has yet to be performed.

In this paper we  use the Finite-Difference-Time-Domain (FDTD) \cite{taflove}
 method to study the time-dependent properties of  CW propagation
in multilayer structures with a Kerr type nonlinearity. We find our 
results generally in accord to those  obtained for systems with weak 
linear index modulation \cite{desterke}, which were solved with approximate 
methods. 
We next examine the dynamics of driving the system from one transmission 
state to 
the other by injecting a pulse, and try to find the optimal pulse 
parameters for this switching. We also test how these parameters 
change for a different initial phase or frequency of the 
pulse.
Finally, we will repeat all work for the case of a linear multilayer
structure with a nonlinear impurity layer.

\section{FORMULATION} 

Electromagnetic wave propagation in dielectric media is governed by
Maxwell's equations
\begin{eqnarray}
\mu \frac{\partial \vec{H}}{\partial t}  = 
 -\vec{\nabla} \times \vec{E} & \hspace{1.5cm} &  
\frac{\partial \vec{D}}{\partial t}  = 
\vec{\nabla} \times \vec{H}
\end{eqnarray}
Assuming here a Kerr type saturable nonlinearity and an isotropic medium, 
the electric flux density
$\vec{D}$ is related to the electric field $\vec{E}$ by
\begin{eqnarray}
\vec{D}=\epsilon_0 \vec{E} +\vec{P}_L +\vec{P}_{NL}=\epsilon_0
\left( \epsilon_r +\frac{\alpha 
\vert \vec{E} \vert^2}{1+\gamma \vert \vec{E} \vert^2}\right) \vec{E}
\end{eqnarray}
where $\gamma \geq 0$.
$\vec{P}_L$ and $\vec{P}_{NL}$ are the induced linear and nonlinear 
electric polarizations
respectively. Here we will assume zero linear dispersion and so a 
frequency independent $\epsilon_r$.
Inverting this to obtain $\vec{E}$ from $\vec{D}$ involves the 
solution of a cubic equation in $\vert \vec{E} \vert$. For $\alpha \geq 0$
there is always only one real root, so there is no ambiguity. For 
$\alpha < 0$ this is true only for $\gamma > 0$. In our study we
will use $\alpha =-1$ and $\gamma = (\epsilon_r-1)^{-1} $ so that 
for $\vert \vec{E} \vert \rightarrow \infty$, 
$\vec{D}\rightarrow\epsilon_0\vec{E}$.

The structure we are considering consists of a periodic array of 21
nonlinear dielectric layers in vacuum, each 20 nm wide with 
$\epsilon_r=3.5$, separated by a lattice constant $a=$200 nm. The linear, 
or low intensity,
transmission coefficient as a function of frequency is shown in 
Fig.~1a. In the numerical setting each unit cell is divided into 256 grids, 
half of them defining the highly refractive nonlinear layer. For the midgap
frequency, this corresponds to about 316 grids per wavelength in the
vacuum area and 1520 grids per effective wavelength in the nonlinear 
dielectric, where of course the length scale is different in the two regions.
 Stability considerations only require more than 20 grids
per wavelength \cite{taflove}. Varying the number of the grids used we found 
our results
to be completely converged. On the two sides of the system we
apply absorbing boundary conditions \cite{taflove}. 

We first study the structure's response to an incoming constant
plane wave of frequency
close to the gap edge $\omega a/2\pi c=0.407$. For each value of the 
amplitude, we wait until the system reaches a steady state and then 
calculate the corresponding transmission and reflection coefficients.
If no steady state is achievable, we approximate them by averaging
the energy transmitted and reflected over a certain period of time, 
always checking that energy conservation is satisfied. Then the
incident amplitude is increased to its next value, which is done 
adiabatically over a time period of 20 wave cycles, and measurements
are repeated. This procedure continues until a desired maximum 
value is reached, and then start decreasing the amplitude, repeating 
backwards the same routine. The form of the incident CW is
\begin{eqnarray}
E_{\tiny{CW}}(t)=\left(A_{\tiny{CW}}+dA_{\tiny{CW}}
\frac{min \{ (t-t_0),20T \} }{20T}\right)e^{i\omega t}
\end{eqnarray}
where $t_0$ is the time when the amplitude change started, $A_{\tiny{CW}}$ is 
the last amplitude value considered and $dA_{\tiny{CW}}$ the amplitude 
increment. 
One wave cycle $T$  involves about 2000 time steps.

\section{RESPONSE TO A CW BIAS}

The amplitude  of the CW is varied from zero to a maximum of 0.7 
with about 40 measurements in between. Results are shown in Fig.~1b,
along with the corresponding one from a time-independent approximation.
The agreement between the two methods is exact for small intensities,
however,  after  a certain input the output waves are not constant any
more but pulsative. This is in accord with the results
obtained with the slowly varying envelope approximation for systems
with a weak refractive index modulation \cite{desterke}. It is interesting
 that the averaged output power is still in agreement with
the time-independent results, something not mentioned in earlier work.
For higher input values, the solution will again reach a steady state just
before going to the second nonlinear jump, after which it will 
again become pulsative. This time though, the averaged transmitted power 
is quantitatively different from the one predicted from time-independent
calculations. 

The nonlinear transmission jump originates from the excitation of a 
stationary gap soliton  when the incident intensity 
exceeds a certain threshold value. Due to the nonlinear change of the 
dielectric constant, the photonic gap is shifted locally in the area 
underneath the soliton, which becomes  effectively 
transparent,  resembling a quantum well with  the soliton being
its bound state solution \cite{lidorikis1}.
The incident radiation coupled to that soliton 
tunnels through the structure and large transmission is achieved. 
We obtain a maximum switching time of the order of 100 $T_r$, or 
a frequency of 360 GHz, where $T_r=2L/c$ is roundtrip time in vacuum.
The second transmission jump is related to the excitation of
two gap solitons, which however, are not stable and so transmission
is pulsative. The Fourier transform of the output shows that,
after the second transmission jump, the system pulsates at a frequency
$\omega a/2\pi c=0.407\pm n \times 0.024$, 
exactly three times the one of the first pulsating solution 
 $\omega a/2\pi c=0.407\pm n \times 0.008$, where $n$ is an integer. 
For much higher input values the response eventually becomes chaotic.
A more detailed description of the switching process as well as the
soliton generation dynamics can be found in \cite{desterke}.

\section{PULSE DRIVEN SWITCHING}

We next turn to the basic objective of this work. We assume a 
specific constant input amplitude  $\vert A_{\tiny{CW}}\vert=0.185$ 
corresponding to  the middle of 
the first bistable 
loop. Depending on the system's history, it can be either in
the low transmission state I, shown in Fig.~1c, or in the high 
transmission state II, shown in Fig.~1d, which are both steady states. 
We want to study the dynamics of a pulse
injected into a system like that. More specifically, if it will
drive the system to switch from one state to the other, how the 
fields change in the structure during switching, for which
pulse parameters this will happen and if these parameters change
for small phase and frequency fluctuations. We assume Gaussian envelope
pulses
\begin{eqnarray}
E_{\tiny{P}}(0,t)=A_{\tiny{P}}e^{-(t-t_0-5t_w)^2/t_w^2}
e^{i\omega t}
\end{eqnarray}
where $A_{\tiny{P}}$ is the pulse amplitude, $t_0$ the time when injection 
starts,
and $W=2t_wc$ is the pulse's full width at  $1/e$ of maximum amplitude.
The beginning of time $t$ is the same as for the CW, so there is
no phase difference between them. 
After injection we wait until the system reaches a steady state again
and then measure the transmission and reflection coefficients to 
determine the final state. During this time we save the field 
values inside the structure every few time steps, as well as the
transmitted and reflected waves. This procedure is repeated for 
various values of $A_{\tiny{P}}$ and $t_w$, for both possible
initial states. Our results are summarized in Fig.~2. White areas 
indicate the pulse parameters for which the intended switch was
successful while black are for which it failed. In Fig.~2a, or
the ``Switch'' graph, the intended switching scheme is for the same
pulse to be able to drive the system from state I to state II
and {\em vise versa}. Fig.~2b, or ``Switch All Up'', is for a pulse
 able to drive the system from I to II, but fails to do 
the opposite, ie. the final state is always II independently from
which the initial state was. Similarly, Fig.~2c or ``Switch All Down'',
is for the pulse whose final state is always I, and Fig.~2d or
''No Switch'', for the pulse that does not induce any switch for 
any initial state.

We find  a rich structure on these parameter planes. Note also that
there is a specific cyclic order as one crosses the curves moving to
higher pulse energies: $\rightarrow$ d $\rightarrow$ b $\rightarrow$ 
a $\rightarrow$ c $\rightarrow$ etc. This indicates that there must
be some kind of energy requirements for each desired switching scheme.
After analyzing the curves it was found that only the first one
in the ``Switch'' graph could be assigned to a simple constant energy
curve $\mathcal{E}$$\sim W \vert A_p\vert^2$. Since any switching involves
the creation or destraction of a stationary soliton, then this
should be its energy. In order to put some numbers, if we would assume
a nonlinearity $\vert \alpha\vert=10^{-9}$ cm$^2$/W, then we would need
a CW of energy $\simeq 34$ MW/cm$^2$ and a pulse of width 
$W/c$ of a few tenths of femtoseconds  and energy $\mathcal{E}$$\simeq 
2.5 \mu$J/cm$^2$. These energies may seem large, but they
can be sufficiently lowered by increasing the number of layers 
and using an incident frequency closer the the gap edge.

In order to find more about how the switching occurs, we plotted
in Fig.~3 the effective transparent areas and the output fields
as a function of time, for the first three curves of Fig.~2a.
As expected, for the pulse from the first curve, the energy for the
soliton excitation is just right, and the output fields are small
compared to the input. For the other curves however, there is an
excess of energy. The system has to radiate this energy away before
a stable gap soliton can be created. It is interesting to note that 
this energy goes only in the transmitted wave, not the reflected,
and it consists of a series of pulses \cite{eggleton}. For the second curve 
in Fig.~2a 
there is one pulse, for the third there are two etc. The width
and frequency of the pulses are independent of the incident pulse,
they are the known pulsating solutions we found in the CW case. So the 
system temporally goes into a pulsating state to radiate away 
the energy excess before settling down into a stable state. If this
energy excess is approximately equal to an integer number of
pulses (the solitary waves from the unstable solutions), then we will have
a successful switch, otherwise it will fail. A similar behavior is
found in  the system's response during switch down 
for the first three curves in Fig.~2a, using
exactly the same pulses as before for the switch up. So the same pulse is
capable of switching the system up, and if reused, switching the system
back down. Using the numbers assumed before for the nonlinearity $\alpha$, the pulses
used in Fig.~3 are 
(a) $W/c=14$ fs, $\mathcal{E}$=2.5 $\mu$J/cm$^2$,
(b) $W/c=28$ fs, $\mathcal{E}$=12 $\mu$J/cm$^2$,
(c) $W/c=42$ fs, $\mathcal{E}$=32 $\mu$J/cm$^2$.

Up to now, the injected pulse has been treated only as an amplitude
modulation of the CW source, ie. they had  the same exactly  frequency 
and there was no phase difference
between them. The naturally rising question is how an initial
random phase between the CW and the pulse, or a slightly different 
frequency affect our results. We repeated the simulations for
various values of  an initial phase difference, first
keeping them with the same frequency. We find that although the results 
show  qualitatively the
same stripped structure as in Fig.~2, there are  quantitative differences. 
The main result is that there
is not a set of pulse parameters that would perform the desired
switching successfully for any initial phase difference. 
Thus the pulse can not be incoherent with the CW, ie. generated at different 
sources, if a controlled and reproducible switching mechanism is desired,
but rather it should  be introduced as an amplitude modulation of the CW. 
However, if this phase could be controlled, then the switching operation
would be controlled, and a single pulse would be able to perform all
different operations.

The picture does not change if we use pulses
of slightly different frequency from the source. We used various pulses
with frequencies both higher and lower than the CW, and we  found a 
sensitive, rather chaotic,  dependence on the initial phase at 
injection time. The origin of this complex response,  if it is an 
artifact of the simple Kerr-type nonlinearity model that we used, 
and if it should appear for other kinds of nonlinearities, 
 is not yet clear to as.
More work is also needed on how these results would change if one
used a different $\vert A_{\tiny{CW}}\vert$ not in the middle of the
bistable loop, a wider or narrower bistable loop etc., but these
would go more into the scope of engineering.

\section{LINEAR LATTICE WITH A NONLINEAR IMPURITY LAYER}

Besides increasing the number of layers to achieve lower switching
thresholds, one can use a periodic array of linear layers $\epsilon
=\epsilon_0\epsilon_r$  with a nonlinear impurity layer 
\cite{radic,hattori,wang,lidorikis2}
$\epsilon=\epsilon_0(\epsilon_r^{\prime}+\alpha 
\vert \vec{E} \vert^2)$ where $\epsilon_r \neq \epsilon_r^{\prime}$
and we will use $\alpha=+1$ and $\gamma=0$.
This system is effectively a Fabry-Perot cavity
with the  impurity (cavity) mode inside the photonic gap, as shown in
Fig.~4a.  The bistable response originates from its  nonlinear 
modulation with light intensity. The deeper this mode is in the
gap, the stronger the linear dispersion for frequencies close to it.
Because of the high $Q$ of the mode, we can use frequencies extremely
close to it achieving very low switching thresholds \cite{lidorikis2}.
 Here however we only want
to study the switching mechanism, so we will use a shallow impurity mode.

The bistable input-output diagram,
the output fields during switching and the field distributions
in the two transmission branches are also shown in Fig.~4. We observe 
a smaller relaxation time and of course the absence of pulsating solutions.
 The parameters
used are $\epsilon_r^{\prime}=1$ and $\omega a/2 \pi c=0.407$ which
corresponds to a frequency between the mode and the gap edge. 
We want to test if a pulse can drive this system in switching 
between the two different transmission states, and again test our results
against phase and frequency perturbations. The two states shown in Fig.~4
are for an input CW amplitude of $\vert A_{\tiny{CW}}\vert=0.16$.
The results for coherent, pulse and CW, are shown in Fig.5. We see that
any desired form of switching can still be achieved, but the 
parameter plane graphs no more bare any simple explanations as the 
ones obtained for the nonlinear superlattice.
Repeating the simulations for incoherent beams and different
frequencies we obtain the same exactly results as before. Only
phase-locked beams can produce controlled and reproducible
switching.

\section{CONCLUSIONS}

We have studied the time-dependent switching properties 
of nonlinear dielectric multilayer systems for frequencies
inside the photonic band gap of the corresponding linear structure. 
The system's response is characterized by both stable and self-pulsing
solutions. We examined the dynamics of driving the system between
different transmission states by pulse injection, and found correlations
between the pulse, the stationary gap soliton and the unstable
solitary waves. A small dependence on the phase difference
between the pulse and the CW is also found, requiring coherent
beams for fully controlled and reproducible
switching. Similar results are also found for the case of a
linear periodic structure with a nonlinear impurity.

\acknowledgements 
Ames Laboratory is operated for the U. S. Department of Energy by Iowa 
State University under contract No. W-7405-ENG-82. This work was supported
by the Director of Energy Research office of Basic Energy Science and 
Advanced Energy Projects, the Army Research office, and a PENED grand.

\begin{figure}
\psfig{figure=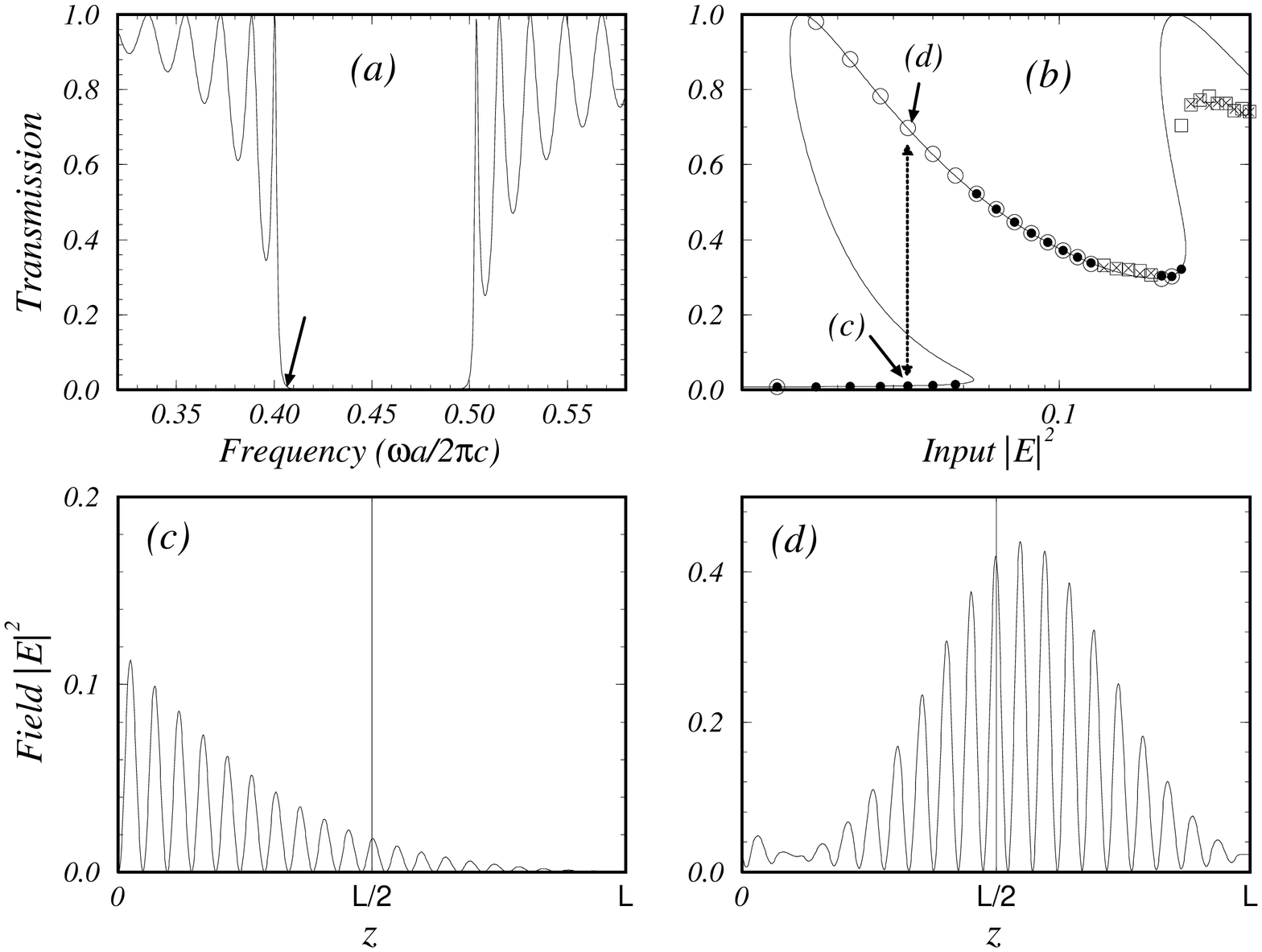,width=16cm,height=16cm,angle=270}
\caption{(a) The linear transmission diagram. The
small arrow indicates the frequency we used in the nonlinear study.
(b) The nonlinear response: closed/(open) circles  correspond
to steady states when increasing/(decreasing) the intensity and
crosses/(open squares) correspond to self-pulsing states when 
increasing/(decreasing) the intensity.
(c) Intensity configuration for low transmission state.
(d) Intensity configuration for high transmission state. }
\end{figure}



\begin{figure}
\psfig{figure=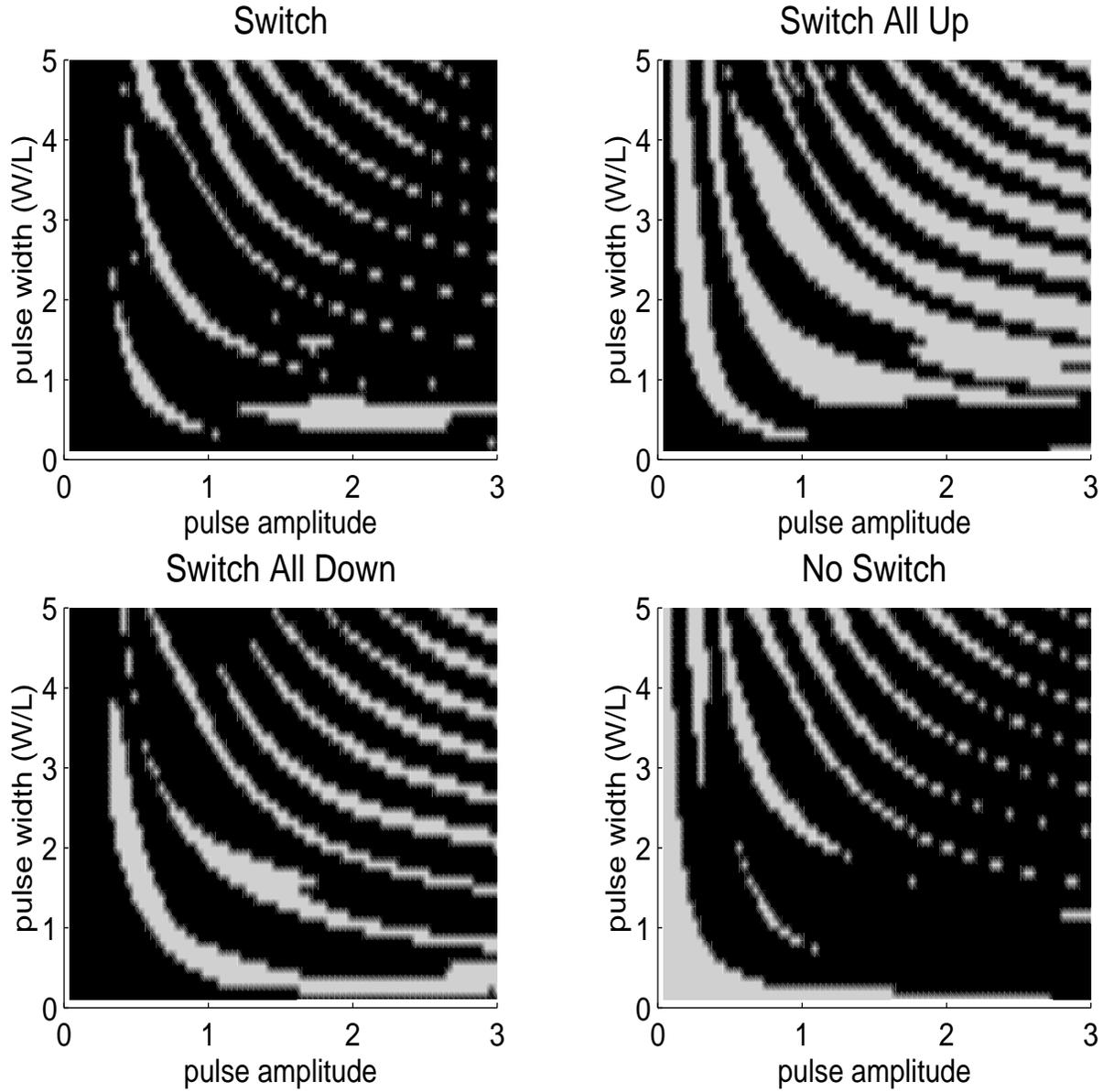,width=16cm,height=16cm,angle=0}
\caption{The four different switching schemes: 
(a) Final state opposite of initial,
(b) final state always high transmission state,
(c) final state always low transmission state,
(d) no change of state. White areas indicate successful 
operation while  black indicate failure.
}
\end{figure}

\begin{figure}
\psfig{figure=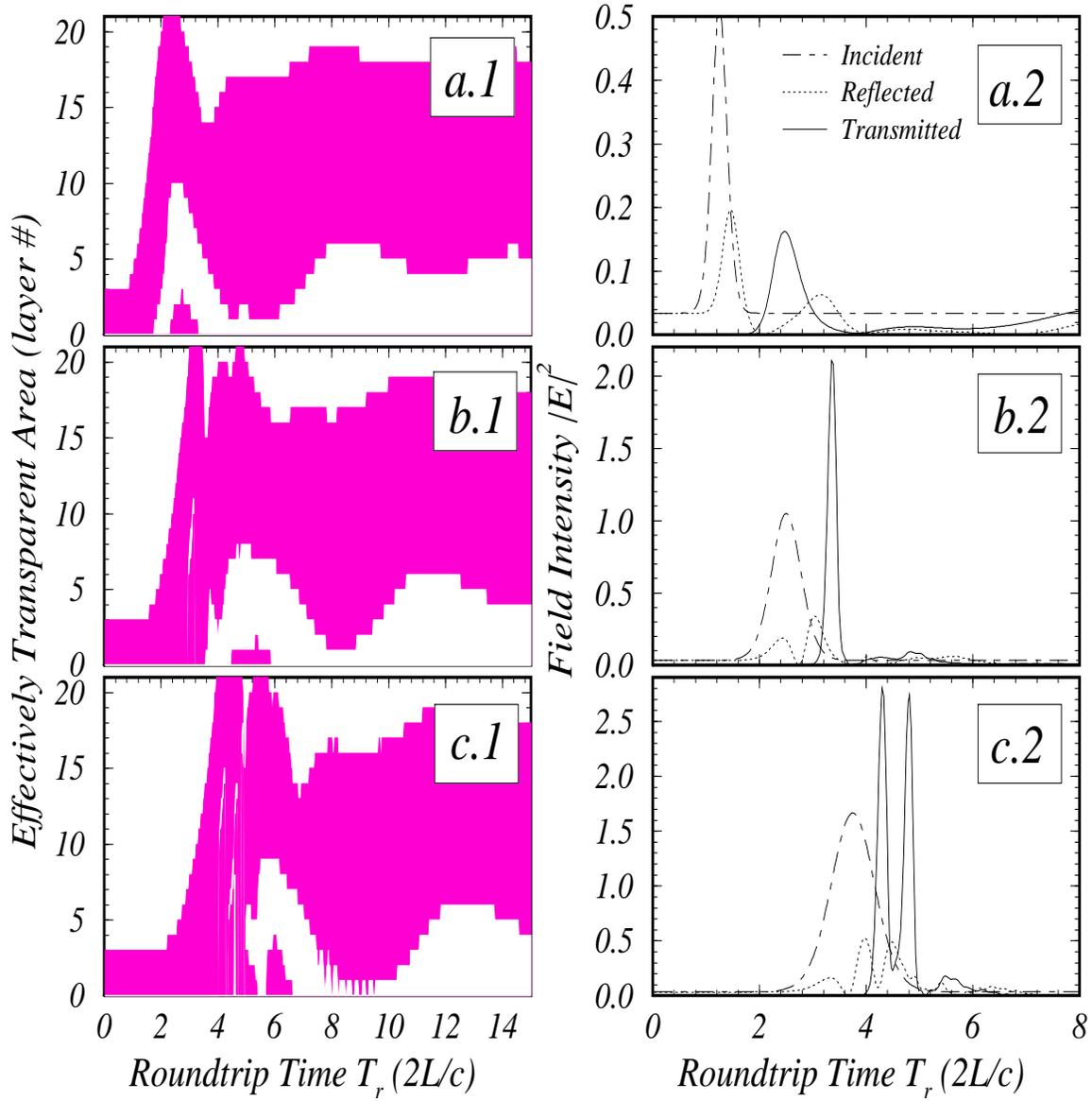,width=16cm,height=16cm,angle=270}
\caption{Switch-up dynamics for (a) First, (b) second and 
(c) third white curves in the ``Switch'' graph of Fig.~2.
(1) Effectively transparent area and (2) input and output waves.}
\end{figure}



\begin{figure}
\psfig{figure=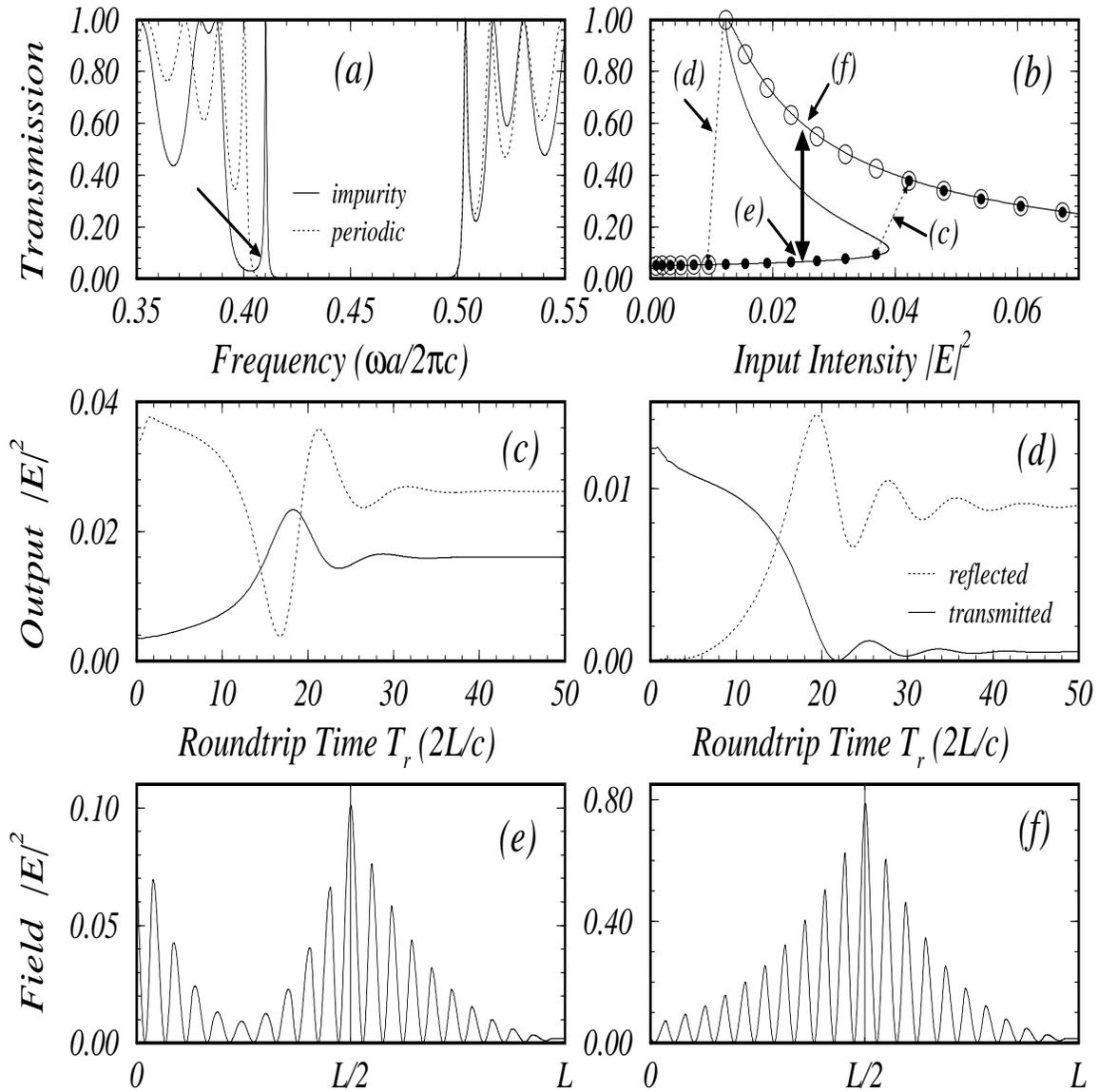,width=16cm,height=16cm,angle=270}
\caption{Linear lattice with a nonlinear impurity layer: 
(a) Linear transmission diagrams, (b) nonlinear
response; solid/(open) circles for increasing/(decreasing)
CW intensity, (c) output waves during switch-up and 
(d) switch-down, (e) intensity configuration for the low transmission
state and (f) high transmission state.}
\end{figure}

\begin{figure}
\psfig{figure=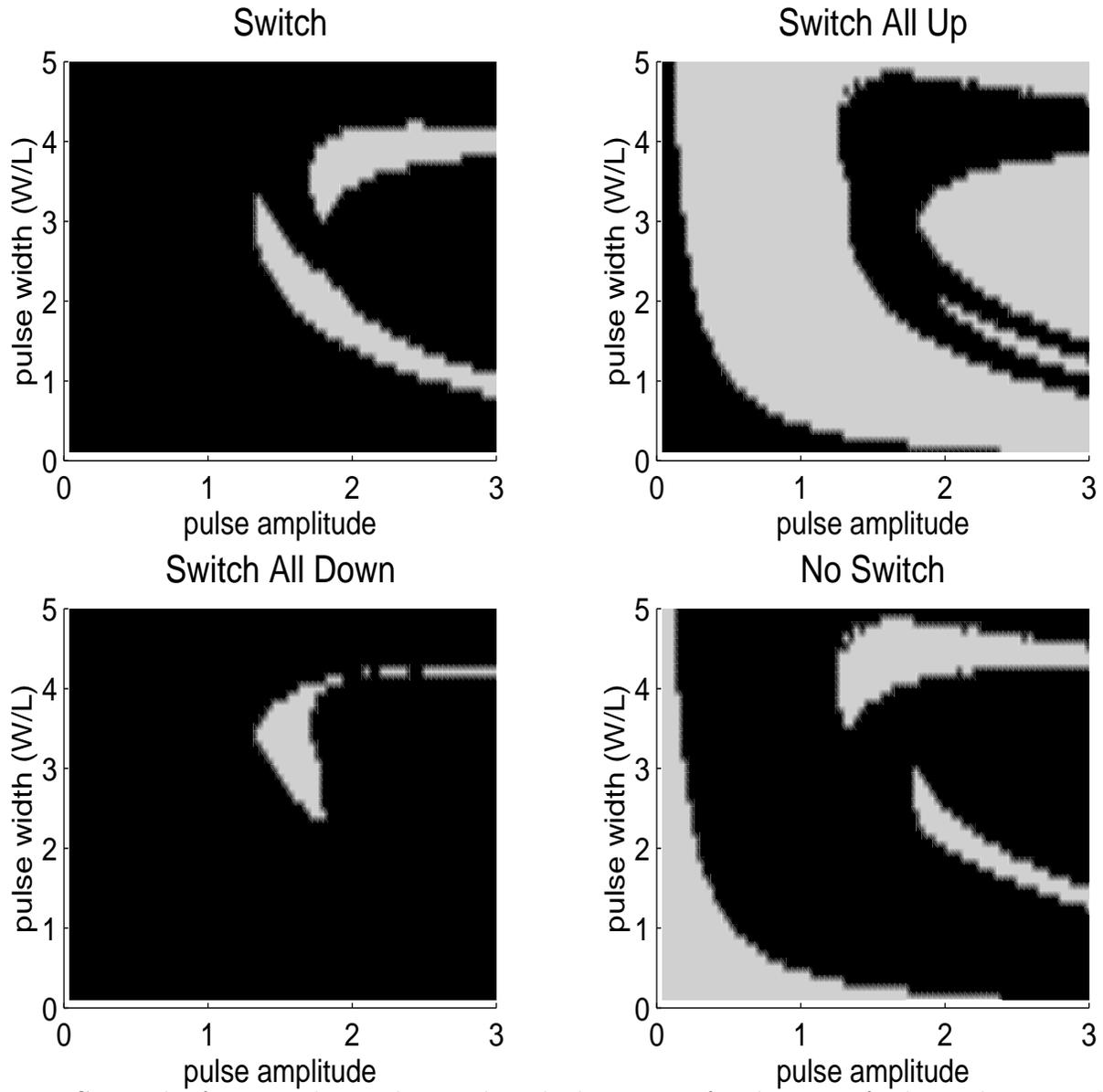,width=16cm,height=16cm,angle=0}
\caption{The four switching schemes described in Fig.~2 for the
case of a linear lattice with a nonlinear impurity layer.
No simple curved structure is found here. }
\end{figure}

\end{document}